\documentstyle[12pt]{article}

\catcode`\@=11
\long\def\@makefntext#1{
\protect\noindent \hbox to 3.2pt {\hskip-.9pt  
$^{{\ninerm\@thefnmark}}$\hfil}#1\hfill}		

\def\@makefnmark{\hbox to 0pt{$^{\@thefnmark}$\hss}}  
	
\def\ps@myheadings{\let\@mkboth\@gobbletwo
\def\@oddhead{\hbox{}
\rightmark\hfil\ninerm\thepage}   
\def\@oddfoot{}\def\@evenhead{\ninerm\thepage\hfil
\leftmark\hbox{}}\def\@evenfoot{}
\def\sectionmark##1{}\def\subsectionmark##1{}}

\renewcommand{\thefootnote}{\fnsymbol{footnote}}

\newcounter{sectionc}\newcounter{subsectionc}\newcounter{subsubsectionc}
\renewcommand{\section}[1] {\vspace*{0.6cm}\addtocounter{sectionc}{1} 
\setcounter{subsectionc}{0}\setcounter{subsubsectionc}{0}\noindent 
	{\normalsize\bf\thesectionc. #1}\par\vspace*{0.4cm}}
\renewcommand{\subsection}[1] {\vspace*{0.6cm}\addtocounter{subsectionc}{1} 
	\setcounter{subsubsectionc}{0}\noindent 
	{\normalsize\it\thesectionc.\thesubsectionc. #1}\par\vspace*{0.4cm}}
\renewcommand{\subsubsection}[1]
{\vspace*{0.6cm}\addtocounter{subsubsectionc}{1}
	\noindent {\normalsize\rm\thesectionc.\thesubsectionc.\thesubsubsectionc. 
	#1}\par\vspace*{0.4cm}}

\newcounter{appendixc}
\newcounter{subappendixc}[appendixc]
\newcounter{subsubappendixc}[subappendixc]

\renewcommand{\appendix}[1] {\vspace*{0.6cm}
        \refstepcounter{appendixc}
        \setcounter{figure}{0}
        \setcounter{table}{0}
        \setcounter{equation}{0}
        \renewcommand{\thefigure}{\Alph{appendixc}.\arabic{figure}}
        \renewcommand{\thetable}{\Alph{appendixc}.\arabic{table}}
        \renewcommand{\theappendixc}{\Alph{appendixc}}
        \renewcommand{\theequation}{\Alph{appendixc}.\arabic{equation}}
        \noindent{\bf Appendix \theappendixc #1}\par\vspace*{0.4cm}}

\def\abstracts#1{{
	\centering{\begin{minipage}{12.2truecm}\footnotesize\baselineskip=12pt\noindent
	\centerline{\footnotesize ABSTRACT}\vspace*{0.3cm}
	\parindent=0pt #1
	\end{minipage}}\par}} 

\renewenvironment{thebibliography}[1]
	{\begin{list}{\arabic{enumi}.}
	{\usecounter{enumi}\setlength{\parsep}{0pt}
\setlength{\leftmargin 1.25cm}{\rightmargin 0pt}
	 \setlength{\itemsep}{0pt} \settowidth
	{\labelwidth}{#1.}\sloppy}}{\end{list}}

\newcounter{itemlistc}
\newcounter{romanlistc}
\newcounter{alphlistc}
\newcounter{arabiclistc}

\newcommand{\fcaption}[1]{
        \refstepcounter{figure}
        \setbox\@tempboxa = \hbox{\footnotesize Fig.~\thefigure. #1}
        \ifdim \wd\@tempboxa > 6in
           {\begin{center}
        \parbox{6in}{\footnotesize\baselineskip=12pt Fig.~\thefigure. #1}
            \end{center}}
        \else
             {\begin{center}
             {\footnotesize Fig.~\thefigure. #1}
              \end{center}}
        \fi}

\newcommand{\tcaption}[1]{
        \refstepcounter{table}
        \setbox\@tempboxa = \hbox{\footnotesize Table~\thetable. #1}
        \ifdim \wd\@tempboxa > 6in
           {\begin{center}
        \parbox{6in}{\footnotesize\baselineskip=12pt Table~\thetable. #1}
            \end{center}}
        \else
             {\begin{center}
             {\footnotesize Table~\thetable. #1}
              \end{center}}
        \fi}

\def\@citex[#1]#2{\if@filesw\immediate\write\@auxout
	{\string\citation{#2}}\fi
\def\@citea{}\@cite{\@for\@citeb:=#2\do
	{\@citea\def\@citea{,}\@ifundefined
	{b@\@citeb}{{\bf ?}\@warning
	{Citation `\@citeb' on page \thepage \space undefined}}
	{\csname b@\@citeb\endcsname}}}{#1}}

\newif\if@cghi
\def\cite{\@cghitrue\@ifnextchar [{\@tempswatrue
	\@citex}{\@tempswafalse\@citex[]}}
\def\citelow{\@cghifalse\@ifnextchar [{\@tempswatrue
	\@citex}{\@tempswafalse\@citex[]}}
\def\@cite#1#2{{$\null^{#1}$\if@tempswa\typeout
	{IJCGA warning: optional citation argument 
	ignored: `#2'} \fi}}

 1
 1
 1

\font\ninerm=cmr9

\begin{document}

\centerline{\normalsize\bf STERILE NEUTRINOS: PHENOMENOLOGY AND THEORY}
\baselineskip=22pt

\centerline{\footnotesize RABINDRA N. MOHAPATRA}
\baselineskip=13pt
\centerline{\footnotesize\it Department of Physics, University of Maryland}
\baselineskip=12pt
\centerline{\footnotesize\it College Park, MD 20742, USA}
\centerline{\footnotesize E-mail: rmohapat@physics.umd.edu}

\vspace*{0.9cm}
\abstracts{A sterile neutrino in addition to the three known neutrinos 
seems unavoidable if one wants a simultaneous understanding of the 
results of the LSND experiment indicating $\nu_{\mu}-\nu_e$ oscillation 
together with other evidences for
neutrino oscillations such as those from solar and atmospheric neutrino 
deficits. A major theoretical challenge then is to understand why the
sterile neutrino is so light.
A simple solution appears to be to assume that it is 
the lightest neutrino of a mirror sector of the universe which has 
identical matter and gauge content as the standard model. After a brief 
review of the phenomenology, a recently constructed
realistic grand unified model based on the gauge group
$SO(10)\times SO(10)^\prime$ that embeds the mirror universe idea
is presented. Detailed predictions for known and the sterile neutrino 
sector are given and their consequences for cosmology are mentioned.}
 
\normalsize\baselineskip=15pt
\setcounter{footnote}{0}
\renewcommand{\thefootnote}{\alph{footnote}}
\section{Why we may need a sterile neutrino?}

The announcement by the Super-Kamiokande collaboration\cite{sk} of the
evidence for neutrino oscillation (and hence nonzero neutrino mass) in 
their atmospheric neutrino data is a major milestone in 
the search for new physics beyond the standard model, which predicts that 
neutrinos are massless. In addition to the Super-Kamiokande atmospheric 
neutrino data, there are now strong indications for
neutrino oscillations from the five 
solar neutrino experiments (Kamiokande, Homestake, Gallex, Sage and
Super-Kamiokande\cite{expt,superK}), other atmospheric neutrino observations
\cite{atmos} and the direct laboratory observation 
in the LSND experiment\cite{LSND}. To explain all three 
experiments, three 
different scales for mass differences ($\Delta m^2$) are needed.
The atmospheric neutrino data requires $\Delta m^2_{\mu-X}\sim 3\times 
10^{-4}-7\times 10^{-3}$ eV$^2$ whereas the solar neutrino data prefers 
either $3\times 10^{-6}-7\times 10^{-6}$eV$^2$ or $\sim 10^{-10}$ eV$^2$
depending on whether the solution arises via the MSW mechanism or via  
oscillation in vacuum. The LSND data on the other hand prefers
$0.3$eV$^2\leq \Delta m^2_{e\mu}\leq 2.0$ eV$^2$ with $\Delta m^2$ as 
high as 10 eV$^2$ also in the allowed range. Since with the three known 
neutrinos one can get only 
two independent $\Delta m^2$'s, it has been suggested\cite{caldwell} that 
a fourth sterile neutrino be invoked. For the uninitiated a sterile 
neutrino is defined as one whose interaction with standard model 
particles (such as the $W, Z$ etc) is much weaker than the strength of usual 
weak interaction. 
The reason for this is the discovery at LEP and SLC that only three 
neutrino species couple to the Z-boson. It is the goal of this talk to
discuss a possible theoretical scenario for the sterile neutrino after a 
brief discussion of how its introduction solves the neutrino puzzles. 

\section{Scenarios for solving the neutrino puzzles}

In the presence of this extra 
neutrino species ($\nu_{s}$), one can construct several 
scenarios for solving the three neutrino 
puzzles\cite{caldwell,giunti,foot,smirnov}. 
In the original paper introducing the sterile neutrino to solve the 
neutrino puzzles\cite{caldwell}, it was proposed that the solar neutrino 
puzzle is 
solved via the oscillation of the $\nu_e$ to $\nu_s$ using the MSW 
mechanism\cite{msw} and the atmospheric neutrino puzzle is solved 
via the $\nu_{\mu}-\nu_{\tau}$ oscillation with maximal mixing. The solar 
neutrino puzzle 
fixes the $\Delta m^2_{e-s}\simeq (0.35-.75)\times 10^{-5}$ eV$^2$, 
whereas the atmospheric neutrino puzzle implies that $\Delta 
m^2_{\mu-\tau}\simeq 10^{-3}$ eV$^2$. This gives a picture where the 
$\nu_s$ and $\nu_e$ are close by in mass with nearly zero mass and the 
$\nu_{\mu}$ and $\nu_{\tau}$ are nearly degenerate in mass. The masses of 
the $\nu_{\mu}-\nu_{\tau}$ system is determined by the LSND experiment 
 $m_{\nu_2}\simeq m_{\nu_3}\simeq\sqrt{ \Delta m^2_{LSND}}$. If the 
universe has a hot component in its dark matter, as some recent analyses 
suggest\cite{silk}, then this requires the $\nu_{\mu}$ and $\nu_{\tau}$ 
masses to be each 2-3 eV implying that in LSND and KARMEN\cite{drexlin} one 
should observe a $\Delta m^2\simeq 4-9$ eV$^2$.

This scenarion is testable in the SNO\cite{SNO} experiment once they 
measure the solar neutrino flux ($\Phi^{NC}_{\nu}$) in their neutral current 
data and compare with the corresponding charged current value 
($\Phi^{CC}_{\nu}$). If the solar neutrinos convert to active neutrinos, 
then one would expect $\Phi^{CC}_{\nu}/\Phi^{NC}_{\nu}\simeq .5$, whereas 
in the case of conversion to sterile neutrinos, the above ratio would be 
nearly $\simeq 1$.

A second scenario advocated in Ref.\cite{foot} and in 
\cite{smirnov}
suggests that it is the atmospheric $\nu_{\mu}$'s that oscillate into the
sterile neutrinos whereas the solar neutrino oscillation could be involving
either active or sterile ones depending on how many sterile neutrinos one 
postulates. The present atmospheric neutrino data cannot distinguish 
between the $\nu_{\tau}$ and $\nu_s$ as the final oscillation products. 
There is however an interesting suggestion\cite{vissani} that monitoring the
pion production via the neutral current reaction $\nu_{\tau} + 
N\rightarrow \nu_{\tau} +\pi^0 +N$ (which is absent in the case of 
sterile neutrinos) can help in distinguishing between these two 
possibilities.

A possible mass matrix for the first case is\cite{satya}:
\begin{eqnarray}
M=\pmatrix{\mu_1&\mu_3& 0& 0\cr
             \mu_3& 0 & 0 &\epsilon \cr
             0& 0 &\delta & m\cr
             0 & \epsilon & m &\pm\delta\cr}.
\end{eqnarray}
Solar neutrino data requires $\mu_3\ll \mu_1
\simeq2\times 10^{-3}~eV$ and $\epsilon \simeq .05 m $. In the case with the
negative sign in the 44-entry, the $\Delta m^2$ in the atmospheric data as
well as the mixing in the LSND oscillation are linked to one another.
An interesting prediction of this mass matrix is that barring extreme fine 
tuning of parameters, the effective mass measured in neutrinoless double 
beta decay would have an upper bound $<m_{\nu}>\leq 2\times 10^{-3}$ eV.

Finally, it has recently been pointed out \cite{fuller} that the introduction
of the sterile neutrino seems to alleviate the problems connected with 
understanding of the heavy element (elements beyond $Fe$) nucleosynthesis
by r-processes around the supernovae. This may be taken as an independent
argument for the introduction of sterile neutrinos regardless of the 
situation in the neutrino oscillation observations.

Finally, one has to ensure that all the new light particles introduced 
to explain the sterile neutrino do not spoil the success of big bang 
nucleosynthesis which cannot tolerate more than $1.5$ extra 
neutrinos\cite{sarkar}. The contribution of a sterile neutrino is 
governed by its mass difference -squared and mixing with the normal 
neutrinos. For instance, the contribution of a sterile neutrino is 
suppressed\cite{chizov} as long as the following inequality is satisfied:
\begin{eqnarray}
\Delta m^2 sin^42\theta \leq 3\times 10^{-6}~~eV^2
\end{eqnarray}
Any theoretical model must respect this constraint.

\section{Mirror universe theory of the sterile neutrino}

If the existence of the sterile neutrino becomes 
confirmed say, by a confirmation of the LSND observation of $\nu_{\mu}-\nu_e$
oscillation or directly by SNO neutral current data to come in the
early part of the next century, a key theoretical challenge will be to
construct an underlying theory that embeds the 
sterile neutrino along with the others with appropriate mixing pattern,    
while naturally explaining its ultralightness.

It is clear that if a sterile neutrino was introduced into the standard 
model, the gauge symmetry does not forbid a bare mass for it implying that
there is no reason for the mass to be small. It is a common experience in 
physics that if a particle has mass lighter than normally expected on the
basis of known symmetries, then it is an indication for the existence of new
symmetries. This line of reasoning has been pursued in recent literature
to understand the ultralightness of the sterile neutrino by using new 
symmetries beyond the standard model.

We will focus on the recent suggestion that the ultralightness of the   
$\nu_s$ may be related to the existence of a 
parallel standard model\cite{bere,foot,blini} which is an exact copy of
the known standard model (i.e. all matter and all gauge forces identical).
The mirror sector of the model will then have three light neutrinos
which will not couple to the Z-boson and would not therefore have been 
seen at LEP. We will call the $\nu'_i$ as the sterile neutrinos of which 
we now have three. The lightness of $\nu'_i$ is dictated by the mirror 
$B'-L'$ symmetry in a manner parallel to what happens in the standard model. 
The two ``universes'' communicate only via gravity or other forces that are
equally weak. This leads to a mixing between the neutrinos of the two 
universes and can cause neutrino oscillation between $\nu_e$ of our 
universe to $\nu'_e$ of the parallel one in order to explain for example 
the solar neutrino deficit.

  At an overall level, such a picture emerges quite naturally in 
superstring theories which lead to $E_8\times E_8^\prime$ gauge theories 
below the Planck scale with both $E_8$s connected by gravity. We 
hasten to 
emphasize that despite this apparent promising connection, no vacuum 
state that leads to the details needed for our neutrino model has 
been discussed to date. In this paper, we will assume the sub-Planck GUT 
group to be a subgroup of $E_8\times E^\prime_8$ in the hope of possible 
future string embedding of our model. For alternative theoretical
models for the sterile neutrino, see Ref.\cite{smir}.

As suggested in Ref.\cite{bere}, we will assume that the process of 
spontaneous symmetry breaking introduces asymmetry between the two universes 
e.g. the weak scale $v^\prime_{wk}$ in the 
mirror universe is larger than the weak scale $v_{wk}= 246$ GeV in our 
universe. It was shown in Ref.\cite{bere} 
that with this one assumption alone, the 
gravitationally generated neutrino masses\cite{ellis} can provide a 
resolution of the solar neutrino puzzle
(i.e. one parameter generates both the required $\Delta m^2_{e-s}$ and the
mixing angle $sin^22\theta_{e-s}\simeq 10^{-2}$). 

\section{An $SO(10)\times SO(10)$ realization of the mirror universe idea}

In a recent paper\cite{brahma}, we constructed a complete realistic model 
for known particles and forces and
make detailed numerical predictions for the neutrino sector within a 
grand unified scheme that implements the seesaw mechanism. Since the 
simplest GUT model that implements the seesaw mechanism is based on the 
$SO(10)$, we use SUSY $SO(10)\times SO(10)^\prime$ as our gauge group with 
each $SO(10)$ operating in one sector.

We impose a mirror symmetry\cite{volkas} between the two SO(10) 
sectors so that the field contents as well as the gauge and Yukawa 
couplings in 
the two sectors are identical to each other and all differences between 
them arise from the process of spontaneous symmetry breaking. In 
order to constrain the model further, we impose 
the requirement that it conserve R-parity automatically without using 
any extra global symmetries. This ensures that there is a natural cold 
dark matter candidate in the model. We also impose an additional global
permutation symmetry $S_3$ which plays a key role in ensuring the mass 
degeneracy between the tau and muon neutrinos. 
The connection between the visible and the mirror sector occurs via the 
mixing of the heavy right-handed neutrinos\cite{collie}. 

 The nontrivial nature of the problem arises from the 
fact that in a GUT framework the neutrino couplings are intimately linked 
to the charged fermion couplings and it is by no means obvious that with 
a simple set of Higgs fields one can make the observed hierarchical 
pattern of the charged fermion masses and mixings compatible with the 
apparent non-hierarchical mass and mixing pattern for the neutrinos.

The fermions of each generation are assigned to the ${\bf (16, 1)\oplus (1, 
16^\prime)}$ representation of the gauge group. We denote them by 
$\Psi_{e,\mu,\tau}$ in the visible sector and by corresponding symbols 
with a prime in the mirror sector (as we do for all fields). The $SO(10)$
symmetry is broken down to the left-right symmetric model by the combination
of ${\bf 45\oplus 54}$ representations in each sector. The $SU(2)_R\times 
U(1)_{B-L}$ gauge symmetry in turn is broken by the ${\bf 126 \oplus 
\overline {126}}$ representations and we take three such representations 
(and denote them by $\Delta_{0,1,2} \oplus \overline {\Delta}_{0,1,2}$). The 
role of the these fields is two-fold: first, they guarantee automatic 
R-parity conservation and second, they lead to the see-saw 
suppression for the neutrino masses\cite{seesaw}.

The standard model symmetry is then broken by the {\bf 10}-dim. Higgs 
fields of which we take three $H_{0,1,2}$. As is well-known, the {\bf 
126}-dim. representation contains in it left-handed triplets with $B-L= 
2$. Due to the presence of the {\bf 54}-Higgs field $S$ in the model, 
couplings of type $\overline \Delta \overline \Delta S$ and $H H S$ are 
allowed and they lead to induced $B-L$ breaking vev's $v_L$ which give a 
direct Majorana mass to the neutrinos leading to the so called type II 
see saw formula\cite{goran} written symbolically as
\begin{eqnarray}
m_{\nu}\simeq fv_L - \frac{m^2_{\nu^D}}{fv_R}
\end{eqnarray}
where $v_R$ is the generic vev of the $\nu^c\nu^c$ component of {\bf 126}.
As was shown in \cite{goran}, the detailed minimization of the potential 
in such theories leads to the conclusion that $v_L$ is also suppressed by 
a see saw like formula (i.e. $v_L\simeq v^2_{wk}/\lambda v_R$ where 
$\lambda$ is an unknown parameter in the superpotential). If we choose
$v_R\simeq 10^{14}-10^{15}$ GeV (so that it is not far from the GUT scale)
and $\lambda \simeq 0.1-0.01$, then we get $v_L$ in the eV range. Note that
while the second term in Eq. (3) arising from the conventional see saw 
formula leads to a hierarchical mass pattern for the neutrinos, the 
first term has no such obligation. Thus, if we require some neutrinos to 
be nearly degenerate, the first term has to be given the dominant role
as we do here.

A second point is that in the effective MSSM derived from the model, the
low energy Higgs doublets will be assumed to be linear combinations of
the doublets present in all {\bf 10} as well as {\bf 126} dimensional
multiplets. In principle this situation can be realized by appropriate
arrangement of parameters.

Next we assume the invariance of the action under a discrete permutation 
symmetry $S_3$ under which the $(\Psi_{\mu}, \Psi_{\tau})$, 
$(\Delta_1,\Delta_2)$ and $(H_1,H_2)$ transform as doublets whereas 
$\Psi_e$, $H_0$ $\Delta_0$ and the rest of the fields transform as 
singlets. The
same discrete operates in the mirror sector (i.e. no mirror version of
$S_3$). This then restricts the form of the Yukawa part of the superpotential
to the following form:
\begin{eqnarray} 
 W_Y = h_1 \Psi_e \Psi_e H_0 + h_2~ 
(\Psi_{\mu}\Psi_{\mu} + \Psi_{\tau} \Psi_{\tau})~H_0 
 + h_3~\Psi_e (\Psi_{\mu}~H_1+\Psi_{\tau}~H_2)\nonumber\\
+h_4~[(\Psi_{\mu}~\Psi_{\mu}-\Psi_{\tau}~\Psi_{\tau})~H_1
 +2~\Psi_{\mu}~\Psi_{\tau}~H_2]
+ f_1~\Psi_e~\Psi_e~\overline{\Delta}_0 \nonumber \\+ 
f_2~(\Psi_{\mu}\Psi_{\mu}+\Psi_{\tau}\Psi_{\tau})~\overline{\Delta}_0  
 +f_3~[(\Psi_{\mu}\Psi_{\mu}-\Psi_{\tau}\Psi_{\tau})\overline{\Delta}_1
+2\Psi_{\mu}\Psi_{\tau} \overline{\Delta}_2] \nonumber\\
 + f_4~\Psi_e~(\Psi_{\mu}\overline{\Delta}_1 
+\Psi_{\tau}\overline{\Delta}_2)
\end{eqnarray}
Using Eq. 4, we can write down the quark and lepton mass matrices for 
the visible sector as follows.
\begin{eqnarray}
M^D_{10} = \pmatrix{ c_1 & c_2 & c_3 \cr
              c_2 & c_4+c_5 & c_6 \cr
              c_3 & c_6 & c_4-c_5}
~~~;~~~
M^D_{126} = \pmatrix{d_5 & d_1 & d_2 \cr
                   d_1 & d_3+d_6 & d_4 \cr
                   d_2 & d_4 & d_6-d_3}
\end{eqnarray}   
\begin{eqnarray}
M^U_{10} = \pmatrix{a_1 & a_2 & a_3 \cr
                  a_2 & a_4+a_5 & a_6 \cr
                  a_3 & a_6 & a_4-a_5}
~~~;~~~
M^U_{126} = \pmatrix{b_5 & b_1 & b_2 \cr
             b_1 & b_3+b_6 & b_4 \cr
             b_2 & b_4 & b_6-b_3}
\end{eqnarray}
\begin{eqnarray}
M^U = M^U_{10}+M^U_{126}~~~;~~~M^D = M^D_{10}+M^D_{126}~~~;\nonumber\\      
~~~M^\nu = M^U_{10} - 3 M^U_{126}~~~;~~~M^L = M^D_{10} - 3 M^D_{126}
\end{eqnarray}
Even though apriori, it may appear from Eq. 5 and 6 that there are 24 
parameters in the visible sector mass matrices, the actual number is 16 
due to the $S_3$ invariance of the theory which yields eight relations 
among them. They are  
$\frac{c_3}{c_6}=\frac{c_2}{c_5}=\frac{a_2}{a_5}=\frac{a_3}{a_6} $; 
$\frac{a_1}{a_4} =\frac{c_1}{c_4}$; 
$\frac{b_1}{b_3}=\frac{d_1}{d_3}=\frac{b_2}{b_4}=\frac{d_2}{d_4}$ and
$\frac{b_5}{b_6}=\frac{d_5}{d_6}$.
We now proceed to determine the remaining parameters in such a way
they give rise to observed fermion masses and quark mixings and vanishing
charged lepton mixings. The best fit is obtained for the following values
for standard model parameters at the GUT scale
 (all masses in GeV units): $m_t=112.00, m_c=0.370,
m_u=0.0011$; $ m_b=1.115, m_s = 0.0148, m_d=0.0013$; CKM mixing parameters
$s_{12}=-0.2201; s_{23}=0.031; s_{13}=0.0039 $ and the lepton masses
$m_e=0.00033, m_{\mu}=0.0699, m_{\tau}=1.1817$ we find the values
for the parameters $a,b,c,$ and $d$ listed in table I. They in turn enable 
us to determine the Dirac neutrino mass matrix for
the visible as well as the mirror sector. Although the Dirac mass 
matrix for the neutrino does not play a significant role in the masses
and mixings in the individual sector, we will see in the next section that 
it plays a crucial role in the mixing between the two sectors.

\section{Connecting the two sectors and predictions for neutrinos}

The neutrino mass matrix has two contributions as is seen from Eq. 3. 
For $v_R$ near $10^{15}$ GeV, the largest entry from the second term in
Eq. 1 is of order $10^{-2}$ eV in the $33$ element and much smaller in 
other places. As far as the first term goes, its form is dictated by Eq. 3
and we choose it as follows:
\begin{eqnarray}
M_{\nu\nu}=\pmatrix{0 & A_l & A^\prime_l \cr
                    A_l & B_l & D_l \cr
                    A^\prime_l & D_l & -B_l}~~{\rm in~eV}
\end{eqnarray}
In order to obtain the predictions for neutrino masses and mixings, we 
need to know the structure of the mass matrices in the mirror sector and 
the connection between the visible and the mirror sector.

The exact mirror symmetry between the visible and the mirror sector implies
that at the level of the superpotential, all couplings in the mirror sector
are identical to those in the visible sector. We will assume that the 
spontaneous symmetry breaking breaks the mirror symmetry so that actual 
mass matrices will exhibit differences. For simplicity, we will assume that 
all doublet vev's in the mirror sector differ by a common ratio from those
in the visible sector (i.e. $v^\prime_{wk}/v_{wk}=\zeta$ ).
Since this asymmetry will effect the fermion masses in the two MSSM's, we
will expect the $B-L$-breaking scales and the GUT scales to be different.
This in turn will imply that the induced triplet vev's will also be different
in the two sectors. Our strategy will therefore be to scale the Dirac mass
matrix for the mirror sector by a common factor but introduce arbitrary
triplet vev's in the mirror sector. 
\begin{eqnarray}
M_{\nu^\prime \nu^\prime }=\pmatrix{\alpha & A_m & A^\prime_m \cr
                       A_m & B_m & D_m \cr
                       A^\prime_m & D_m & -B_m}~~{\rm in~eV}
\end{eqnarray}
Where, $A_m= q_1 A_l$, $A^\prime_m= q_1 A_l$, $D_m= q_2 D_l$ and $B_m=q_2 
B_l$. Let us now try to connect the two sectors which we do by 
postulating the
Higgs fields ${\bf (16, 16^\prime)\oplus (\bar{16}, \bar{16^\prime})}$ 
(denoted by $\chi\oplus \bar{\chi}$). There can now be a connecting term 
between the two sectors given by 
\begin{eqnarray}
W^\prime = g_c\Psi_e\Psi^\prime_e\bar{\chi} 
+g^\prime_c(\Psi_{\mu}\Psi^\prime_{\mu}
+\Psi_{\tau}\Psi^\prime_{\tau})\bar{\chi}
\end{eqnarray}
We now give a vacuum expectation value to the $\nu^c{\nu^c}^\prime$ element
of $\chi\oplus\bar{\chi}$ fields. Then only the right handed neutrinos
of both sectors get connected. This in conjunction with the Dirac masses
of both sectors introduces a mixing matrix between the two sectors of the 
following form (assuming for simplicity $g_c=g^\prime_c$): 
\begin{eqnarray}
M_{\nu\nu^\prime} =g_c M_{\nu^D}M^{-2}_{\nu^c}M_{\nu^D}<\chi> 
\end{eqnarray}
Where,
\begin{eqnarray}
M_{\nu^c \nu^c}=\pmatrix{0          &  A_r  & A^\prime_r \cr
                   A_r    &  B_r  & D_r \cr
               A^\prime_r &  D_r  & -B_r} ~~~{\rm in~GeV}
\end{eqnarray}
Where, $ A_r= l_1 A_l$, $A^\prime_r= l_1 A^\prime_r$, $B_r = l_2 B_l$ 
and $D_r = l_2 D_l$. We diagonalize the complete neutrino mass matrix to 
obtain the following absolute values of the mass eigenvalues (in eV's):
$m_{\nu^\prime_\tau}=90.56, m_{\nu^\prime_\mu}=-90.56, 
m_{\nu^\prime_e}=0.0034$ $ m_{\nu_\tau}=1.51, m_{\nu_\mu}=-1.509, 
m_{\nu_e}=0.001 $. The squared mass diffences (in eV$^2$) are $\Delta 
m^2_{e-s}= 9.9 \times 10^{-6}$, 
$\Delta m^2_{\mu-\tau}=0.003$ and $\Delta m^2_{e-\mu}= 2.27$, where the 
numbers are given in eV$^2$. The fitted values of the parameters are, 
$\alpha=0.005$, $A_l=0.0253$, $A^\prime_l=0.050$, $B_l=0.675$, 
$D_l=1.35$ given in eV 
units, $q_1=10$, $q_2=60$, $l_1=1.2~10^{15}$, $l_2=0.5~10^{15}$ and 
$g_c<\chi>=6.9\times 10^{12}$ given in GeV units. The mixing matrix 
$O^\nu$ of the six neutrinos in the basis ($\nu_e,\nu_{\mu},\nu_{\tau}, 
\nu^\prime_e,\nu^\prime_{\mu},\nu^\prime_{\tau}$) 
is approximately given as,
{\small
\begin{eqnarray} 
O^{\nu} = \pmatrix{-0.99 & 0.037 & 0 & 0.039 & -0.00025 & 0 \cr 
   -0.031 & -0.85 & -0.52 & -0.00072 & 0 & 0 \cr 
   0.019 & 0.525 & -0.85 & -0.00043 & 0 & 0 \cr 
   -0.042 & 0.0014 & 0.00071 & 
    -0.999 & 0.0062 & 0 \cr 
   0 & 0 & 0 & -0.0053 & -0.850 & -0.525 \cr 
   0 & 0 & 0 & 0.0032 & 0.52 & -0.85}
\end{eqnarray}
}
Combining this with the mixing angle for the charged leptons, we obtain 
the final mixing matrix among the four neutrinos which looks identical 
to the corresponding top-left $4\times 4$ submatrix of $O_{\nu}$ with only 
the $\nu_e-\nu_{\tau}$ entry reduced by a factor of 2 because of the 
presence of a small 13 element in the charged leptonic mass matrix. 
We have varied the 
parameters of the model to see the allowed range for the $m_{\nu_{\mu}}$
relevant for the LSND experiment and find consistent solutions for the
range $0.5\leq m_{\nu_{\mu}}/eV \leq 1.5$. Therefore, the $\nu_{\mu}$ and
$\nu_{\tau}$ together could play the role of the hot dark matter for
the upper allowed range of the masses. Also note that the zeros in the
neutrino mixing matrix simply means that those entries are less than
$10^{-4}$. 

We thus see that in this model not only are all three positive 
indications of neutrino oscillations are explained but the mixing between 
the heavier sterile neutrinos $\nu^\prime_{\mu}$ and $\nu^\prime_{\tau}$ 
and the active neotrinos are consistent with all known
oscillation data such as for example the one from the CHOOZ\cite{chooz}. 
What we find very interesting is that with only six parameters describing 
the entire $6\times 6$ neutrino mass matrix (three active and three 
sterile) and every other parameter fixed by the charged 
fermion masses, four neutrino masses and 12 mixing
parameters that link the active to sterile neutrinos which could have 
observable consequences are all completely  consistent with known data.
   
Turning now to the consistency of our model with big bang nucleosynthesis
(BBN), we recall that present observations of Helium and deuterium abundance
can allow for as many as $4.53$ neutrino species\cite{sarkar} if the baryon
to photon ratio is small. In our model, since the neutrinos decouple
above 200 MeV or so, their contribution at the time of nucleosynthesis
is negligible (i.e. they contribute about $0.3$ to $\Delta N_{\nu}$.)
On the other hand the mirror photon could be completely in equilibrium at
$T=1$ MeV so that it will contribute $\Delta N_{\nu}=1.11$. All together
the total contribution is les than $1.5$.

In conclusion, if the LSND result stands the test of time, this
would indicate an ultralight sterile neutrino. An interesting model
for understanding its lightness is to postulate a mirror universe.
This can be converted into a fully grand unified model within the 
$SO(10)\times SO(10)$ group.

I am grateful to D. Caldwell, Z. Berezhiani and B. Brahmachari for 
collaboration and discussions on the sterile neutrino idea.
I am also very grateful to B. Kniehl and G. Raffelt for kind hospitality
at the Ringberg castle. This work is supported by the National Science 
Foundation under grant no. PHY-9802551.

\begin{table}[htb]
\begin{center}
\[
\begin{array}{|c||c||c||c||c||c|}
\hline
a_1&a_2&a_3&a_4&a_5&a_6 \\
0.140&0.230&0&50.54&-55.30&0 \\
\hline
\hline
c_1&c_2&c_3&c_4&c_5&c_6 \\
0.0016&0.0023&-0.00012&0.598&-0.55&0.0288 \\
\hline
b_1&b_2&b_3&b_4&b_5&b_6 \\
0.140 & 0.230 & 0 & 50.54 & -55.30 & 5.64\\
\hline
\hline
d_1&d_2&d_3&d_4&d_5 & d_6\\
0.0016&0.0023&-0.00012&0.598&-0.55 & -0.0152 \\
\hline
\end{array}
\] 
\end{center}
\caption{The fitted values of $a_i$, $b_i$, $c_i$ and $d_i$}
\label{table}
\end{table}

\end{document}

\begin{table}[h]
\tcaption{$\Gamma(K\rightarrow\pi\pi\gamma)$ for the $K^0_S$,
$K^0_L$ and $K^-$ mesons.}\label{tab:exp}
\small
\begin{tabular}{||c|c|c|l||}\hline\hline
{} &{} &{} &{}\\
Meson &$\Gamma(\pi^+\pi^-)\; s^{-1}$ &$\Gamma(\pi^+\pi^-\gamma)\; s^{-1}$ &{}\\
{} &{} &{} &{}\\
\hline
{} &{} &{} &{}\\
$K^0_S$ &$0.769\times 10^{10}$ &$5.46\times 10^7$ 
&\begin{minipage}{2.5in}
No DE observed, nor (IB)-E1 interference, despite large
statistics, for $E^{\ast}_{\gamma}>20 MeV$.
\end{minipage}\\
{} &{} &{} &{}\\
\hline
{} &{} &{} &{}\\
\raise13pt\hbox{$K^0_L$} &\raise13pt\hbox{$3.93\times 10^4$} 
&\raise13pt\hbox{$0.90\times 10^3$}
&\begin{minipage}{2.5in}
DE prominent, exceeding IB over the range of measurement
$20<E^{\ast}_{\gamma}<160 MeV$.
\end{minipage}\\ 
{} &{} &{} &{}\\[-37pt]
{} &{} &(DE $=0.62\times 10^3)$ &{}\\[24pt]
\hline\hline
\end{tabular} 
\end{table}

\section{Acknowledgements}

\section{References}

\begin{thebibliography}{9}
\end{document}